\documentstyle[12pt]{article}
\newcommand{\beq}{\begin{equation}}
\newcommand{\eeq}{\end{equation}}
\newcommand{\id}
 {i\kern.06em\hbox{\raise.25ex\hbox{$/$}\kern-.60em$\partial$}}

\newcommand{\bs}{/\kern-.52em b}
\newcommand{\qs}{/\kern-.52em s}
\newcommand{\D}{{\cal{D}}}
\newcommand{\dv}{\!d^3\!x\,}

\newcommand{\dd}
{\kern.06em\hbox{\raise.25ex\hbox{$/$}\kern-.60em$\partial$}}

\newcommand{\tr}{\mathop{\rm tr}\nolimits}
\begin{document}

\title{ On three dimensional bosonization}
\author{
J.C. Le Guillou$^a$\thanks{Also at {\it Universit\'e de Savoie} and at
{\it Institut Universitaire de France}}\,,
E. Moreno$^b$\thanks{Supported by CUNY Collaborative Incentive Grant 991999},
C. N\'u\~nez$^c$ \\
and\\
F.A. Schaposnik$^c$\thanks{Investigador CICBA, Argentina}
\\
~
\\
{\normalsize\it
$^a$Laboratoire de Physique Th\'eorique ENSLAPP}
\thanks{URA 1436 du CNRS associ\'ee
\`a l'Ecole Normale Sup\'erieure de Lyon et \`a
l'Universit\'e de Savoie}\\
{\normalsize\it
LAPP, B.P. 110, F-74941 Annecy-le-Vieux Cedex, France}\\
~\\
{\normalsize\it
$^b$ Physics Department, City College of the City University
of New York}\\
{\normalsize\it
New York NY 10031, USA}\\
{\normalsize\it
Physics Department, Baruch College, The City University of 
New York}\\
{\normalsize\it
New York NY 10010, USA}
 \\
~\\
{\normalsize\it
$^c$Departamento de F\'\i sica, Universidad Nacional de La Plata}\\
{\normalsize\it
C.C. 67, 1900 La Plata, Argentina}}
\date{}
\maketitle

\vspace{-5.5 in}

\hfill\vbox{
\hbox{~~}
\hbox{\it ENSLAPP-A-641/97}
\hbox{\it CCNY-HEP 97/2}
\hbox{\it LA PLATA-Th 97/01}
}
\vspace{4.5 in}

\begin{abstract}
{We discuss Abelian and non-Abelian
 three dimensional
bosonization  within the path-integral framework.
We present a systematic approach leading to
the construction of the bosonic action which,
together with the bosonization recipe for fermion currents,
describes the original fermion system in terms of 
vector bosons.
 }
\end{abstract}
\newpage
%

Three dimensional bosonization, i.e. the mapping of a three
dimensional fermionic theory onto an equivalent bosonic quantum field
theory, has attracted considerable attention, particularly
after its connection with Chern-Simons (CS) theories was unraveled
\cite{lut}-\cite{MLNS}. In this context, the path-integral framework
has provided a systematic approach to derive bosonization recipes
both in the Abelian and non-Abelian cases
\cite{FS},\cite{FAS}-\cite{LNS},\cite{MLNS}. This approach relies on
the evaluation of the three-dimensional fermion determinant which, in
contrast with the two-dimensional one, cannot be computed in a closed
form. Because of this problem, one cannot give an exact expression for
the bosonic Lagrangian which corresponds to the original free fermionic
one. In contrast, one does obtain an {\it exact} bosonization recipe
for the fermion current \cite{MLNS} :
\beq
\bar \psi^i \gamma_\mu t^a_{ij} \psi^j \to
\pm \frac{i}{8\pi}\varepsilon_{\mu \nu \alpha}
\partial_\nu A^a_\alpha
\label{1}
\eeq
Here $\psi^i $ ($i,j=1,\cdots,N$) are massive fermions (with mass $m$),
$t^a$ are the generators 
of some group G ($a=1,\cdots,{\rm dim}G$). Finally
$A_\mu$ is a vector field taking values in the Lie algebra of $G$
($\mu = 0,1,2$). 

It is the purpose of the present work  to develop a systematic approach
which allows to evaluate the bosonic action accompanying  
recipe (\ref{1})  once an approximation scheme for calculating
the fermion determinant is adopted. 
In this respect, we
show how to construct the bosonic
action order by order in a $1/m$ expansion so that
our results extend 
those in ref.\cite{MLNS} where only the $m\to\infty$ limit was
considered. For the Abelian case we also analyze an alternative
(quadratic) approximation which allows to make contact
with bosonization of massless fermions. 

To begin with, let us compare the path-integral
bosonization approaches in $2$ and $3$ dimensions. In the former case
one can show that the fermion determinant gives, in the non-Abelian
case, the Wess-Zumino-Witten (WZW) action. Now, thanks to the
existence of the Polyakov-Wiegmann identities for WZW
functionals \cite{PW}, the problems posed by path-integrals of
non-qua\-dra\-tic terms can be overcome and the bosonization
recipe can be derived in a very simple way. 

In three dimensions the situation is more involved. Firstly, one has
not a closed expression for the fermion path integral. Even if one
approximates the fermion determinant by its leading order in a $1/m$
expansion (the non-Abelian CS term), one needs at some stage of the
$d=3$  bosonization procedure to decouple a vector field $A_\mu$
(the bosonic counterpart of the original fermionic fields),
identified by the source term, from an auxiliary field $b_\mu$
but no Polyakov-Wiegmann identity is available in $d=3$
to do this work.
Fortunately, 
the existence
of a ``large'' BRST invariance of the
resulting effective action
allows the disentangling of path-integrals
on $A_\mu$ and $b_\mu$. This BRST
invariance was exploited  in \cite{MLNS} in an
analysis of the leading order in the $1/m$ expansion for the
fermion determinant. We show here that in fact it can be
realized independently of any approximation and hence
it ensures that one can always  find the bosonization
recipe for the fermion action.
The BRST symmetry exploited to get the bosonic action  is highly
related to that used in \cite{DNS1}-\cite{DNS5}, and is analogous to
the one that arises in topological field theories  \cite{BS}, its
origin being related to the way an originally ``trivial''  auxiliary
bosonic field enters into play. 

We start by describing the 
main steps in the bosonization technique developed in
\cite{MLNS}.
Consider the $d=3$ Euclidean Lagrangian for $N$ free massive Dirac
fermions
\beq
L = \bar\psi (\id   + m ) \psi
\label{2}
\eeq
The corresponding generating functional reads
\beq
Z_{fer}[s] =  \!\!\int \!\!\D \bar \psi \D \psi
\exp[-\!\!\int d^3x \bar \psi (\id  + \qs + m ) \psi ]
\label{3}
\eeq
where $s_\mu$ is the source for fermion currents. We introduce at
this point an auxiliary  vector field $b_\mu$ through the use of the
$d=3$ identity \cite{MLNS}
\beq
F[s] = \int \!\!Db_\mu
\det(2\varepsilon_{\mu \nu \alpha} D_\nu[b])
\delta( {^*\!\!f}_\mu[b] - {^*\!\!f}_\mu[s])
F[b]
\label{4}
\eeq
where
\beq
^*\!\!f_\mu = \varepsilon_{\mu \nu \alpha}
f_{\nu\alpha} =  
 \varepsilon_{\mu \nu \alpha} (\partial_{\nu} b_{\alpha} 
- \partial_{\alpha} b_{\nu} 
+   [ b_{\nu}, b_{\alpha}])
\label{5}
\eeq
and
\beq
D_\mu[b] = \partial_\mu + [b_\mu,\,\,]
\label{der}
\eeq
Using (\ref{4}) the generating functional (\ref{3}) can be written as
\beq
Z_{fer}[s] =
\int \!\!Db_\mu
\det(2\varepsilon_{\mu \nu \alpha} D_\nu[b])
 \delta( {^*\!\!f}_\mu[b] - {^*\!\!f}_\mu[s])
\det(\id + m + \bs)
\label{6}
\eeq

In order to go on with the bosonization one needs an explicit
expression for the fermion determinant in (\ref{6}). We shall
consider separately the non-Abelian and the Abelian cases. In the
former, we shall extend the results presented in ref.\cite{MLNS} by
discussing the evaluation of the effective action within an expansion
\cite{Red} of the fermion determinant in inverse powers of the
fermion mass. In the latter, apart from briefly discussing the
results of this $1/m$ expansion, we shall also consider an
approximation consisting in keeping quadratic terms in a $b_\mu$
expansion of the fermion determinant \cite{AFZ},\cite{BFO}.
Our analysis will show that even if one does not have
an exact expression for the $d=3$ fermion determinant there is a
systematic way of calculating the bosonic action resulting from the
integration of the auxiliary field $b_\mu$. 

\vspace{0.3cm}
\noindent (i)  {\it The non-Abelian case}
\vspace{0.3 cm}

\noindent Using  identity (\ref{4})  one can  write,
instead of (\ref{6}),
\beq
Z_{fer}[s] = X[s]^{-1}\!\!
\int \!\!Db_\mu X[b]
\det(2\varepsilon_{\mu \nu \alpha} D_\nu[b])
 \delta( {^*\!\!f}_\mu[b] - {^*\!\!f}_\mu[s] )
\det(\id + m + \bs) \label{9}
\eeq
where $X[b]$ is an arbitrary functional which can be introduced in
order to control the issue of symmetries at each stage of the
derivation. It is convenient
to choose  $X$ in the form
\beq
X[b] = \exp(\mp \frac{i}{16\pi} tr \int d^3x b_\mu \,^*\!f_\mu[b] )
\label{car}
\eeq
Now, if we introduce  a Lagrange multiplier $A_\mu$ (taking values
in the Lie algebra of $G$) to represent the delta function
in (\ref{9}), $Z_{fer}[s]$ takes the form
\beq
Z_{fer}[s] = X[s]^{-1}\!\!
\int \D A_\mu \exp \left( \mp \frac{i}{16\pi}
tr \int d^3x  A_\mu  {^*\!\!f}_\mu[s] \right) \times
\exp(-S_{bos}[A])
\label{7}
\eeq
with the bosonic action $S_{bos}[A]$ given by
\begin{eqnarray}
& & \exp(-S_{bos}[A])  =  \int \D b_\mu \det(\id + m + \bs)  \times
\nonumber \\
& & \det(2\varepsilon_{\mu \nu \alpha}D_\nu[b])
\exp\left( \pm \frac{i}{16\pi}
tr \int d^3x  (A_\mu - b_\mu)  {^*\!\!f}_\mu[b] \right)
\label{8}
\end{eqnarray}
Differentiation of $Z_{fer}[s]$ with respect to $s_\mu$ trivially
 leads to the bosonization recipe (\ref{1}) which, we
insist, is an exact result in the sense that no approximation was
done to obtain this result.

The choice (\ref{car})  makes  $A_\mu$  transform
as a gauge connection. Indeed, under the change
\beq
A_\mu \to g^{-1} A_\mu g +  g^{-1} \partial_\mu g
\label{10}
\eeq
\beq
\,b_\mu \to g^{-1} \, b_\mu \,g \,+  g^{-1} \partial_\mu g
\label{11}
\eeq
one has
\beq
S_{bos}[A^g] = S_{bos}[A]
\label{12}
\eeq
which implies the necessary identity
\beq
Z_{fer}[s^g] = Z_{fer}[s]
\label{truc2}
\eeq
This identity guarantees that 
current correlation functions calculated by differentiation of 
the generating functional (\ref{7}) will have 
the correct transformation properties. In this respect,
the bosonization recipe (\ref{1}) 
should  be taken as illustrative of the bosonization since the
rigorous equivalence between the fermionic
and the bosonic theory is at the level of the generating 
functional $Z[s]$ of Green functions. It is from $Z[s]$ 
written in the form (\ref{7}) that
one has to compute 
current correlation functions  in the bosonic language.

We now consider the expression for the bosonic action defined in
eq.(\ref{8}) once ghost fields $\bar c_\mu$ and $c_\mu$ are introduced
to represent the Faddeev-Popov like determinant,
\begin{eqnarray}
\exp(-S_{bos}[A]) & = & \int \D b_\mu \D \bar c_\mu \D c_\mu
\exp \left( -tr \!\int \! \dv ( L[b] \pm \frac{i}{8\pi}
\varepsilon_{\mu \nu \alpha}
\bar c_\mu D_\nu[b] c_\alpha \right.
\nonumber \\
& & \left. \mp \frac{i}{16\pi}
(A_\mu - b_\mu)  {^*\!\!f}_\mu[b] ) \right)
\label{9x}
\end{eqnarray}
The $\pm$ signs are introduced for convenience
and are related to regularization
ambiguities arising in the evaluation of the fermion determinant
which contains a parity violating term with this ambiguity
\cite{Red}. We have written
\beq
tr \!\int \! \dv L[b] = - \log \det (\id + m + \bs)
\label{10x}
\eeq
 It was observed in ref.\cite{MLNS}  that when $L[b]$ is
approximated by its first term in the $1/m$ expansion,
a set of BRST transformations can be defined so
that the corresponding BRST invariance allows to  obtain
the (approximate) bosonic action. We shall 
explicitly prove here that this invariance is present
in (\ref{9x}) where no
approximation for $L[b]$ is assumed.  To this end, we
introduce a set of auxiliary fields $h_\mu$
(taking values in the
Lie algebra of $G$) and
$l$  so that one can rewrite (\ref{9x}) in the
form
\beq
\exp(-S_{bos}[A])   =  \int \D b_\mu \D \bar c_\mu \D c_\mu
\D h_\mu \D l \D \bar \chi \exp (-S_{eff}[A,b,h,l,\bar c, c,
\bar \chi])
\label{cor}
\eeq
with
\begin{eqnarray}
S_{eff}[A,b,h,l,\bar c, c, \bar \chi] & = &
 tr \!\int \! \dv ( L[b-h]
\pm \frac{i}{8\pi}
\varepsilon_{\mu \nu \alpha}
\bar c_\mu D_\nu[b] c_\alpha
\nonumber\\
& &
 \mp \frac{i}{16\pi}
( (A_\mu - b_\mu)  {^*\!\!f}_\mu[b] + l h_\mu^2 - 2 \bar \chi
h_\mu c_\mu ) )  \label{11x}
\end{eqnarray}
where $\bar \chi$ is an anti-ghost field. Written in the form
(\ref{11x}), the bosonic action has a BRST invariance under the
following nilpotent off-shell BRST transformations %
\[
\delta \bar c_\mu = A_\mu - b_\mu , \;\;\;\;\;\;
\delta b_\mu = c_\mu , \;\;\;\;\;\; \delta A_\mu = c_\mu , \;\;\;\;\;\;
\delta c_\mu = 0 , \;\;\;\;\;\;
\delta \bar \chi = l
\]
\beq
\delta h_\mu = c_\mu , \;\;\;\;\;\; \delta l = 0
\label{17}
\eeq
(These transformations are slightly
different from those in ref.\cite{MLNS} due to the present choice of
functional $X[b]$)

In view of this BRST invariance, one could add to $S_{eff}$  a BRST
exact form without changing the dynamics defined by $S_{bos}[A]$.
Exploiting this, 
 we shall see that one can
factor out the $A_\mu$ dependence in
the r.h.s. of eq.(\ref{11x})  so that it completely decouples from
the path-integral over auxiliary and ghost fields. Although
complicated, this integral then becomes irrelevant for the definition
of the bosonic action for $A_\mu$. Indeed, let us add to $S_{eff}$
the BRST exact form $\delta G$,
\beq
S_{eff}[A,b,h,l,\bar c, c,\chi] \to S_{eff}[A,b,h,l,\bar c, c,\chi] +
\delta G[A, b, h, \bar c]
\label{20}
\eeq
with
\beq
G[A, b, h, \bar c] = \mp \frac{i}{16 \pi} tr \!\int \! \dv
\varepsilon_{\mu \nu \alpha} \bar c_\mu H_{\nu \alpha}[A,b,h]
\label{21}
\eeq
and $H_{\nu \alpha}[A,b,h]$ a functional to be determined in order to
produce the decoupling. Then, consider
the change of variables 
\beq
b_\mu = 2b'_\mu - A_\mu + V_\mu[A]
\label{18}
\eeq
where $V_\mu[A]$ is some functional of $A_\mu$ changing covariantly
under gauge transformations,
\beq
V_\mu[A^g] = g^{-1} V_\mu[A] g
\label{19}
\eeq
so that $b'_\mu$ is, like $A_\mu$ and $b_\mu$, a gauge field.
Integrating over $l$ in (\ref{cor}) and imposing the resulting
constraint, $h_\mu = 0$,  one
sees that if one imposes on $H_{\nu \alpha}[A,b,h]$ the condition
\begin{eqnarray}
\varepsilon _{\mu \nu \alpha}\!\!\int\!\!d^3y
 \!\left(
\frac{\delta H_{\nu \alpha}}{\delta  b^a_\rho (y)}\!+
\frac{\delta H_{\nu \alpha }}{\delta A^a_\rho (y)}\!+
\frac{\delta H_{\nu \alpha }}{\delta h^a_\rho (y)}
\right)\!c^a_\rho (y) \left|_{h=0}\right.\!=
\varepsilon _{\mu \nu \rho } [A_\nu\!-\! b_\nu\!-\!V_\nu[A], c_\rho] 
 \nonumber\\
\label{28}
\end{eqnarray}
then, when written in terms of the new $b'_\mu$ variable, the ghost
term becomes
\beq
S_{ghost}[b',c,\bar c] = \pm \frac{i}{8 \pi}
tr \! \int \!\dv \varepsilon_{\mu \nu \alpha}
\bar c_\mu D_\nu[b'] c_\alpha
\label{29}
\eeq
so that its contribution is still $A_\mu$ independent. Then,
we can write the effective action in the form
\beq
S_{eff}[b',A] + S_{ghost}[b',c,\bar c]
\label{truc3}
\eeq
with
\begin{eqnarray}
S_{eff}[b',A] & = & \tilde{S}[b,A]
   \nonumber \\
& = & tr\!\int \! \dv \left( L[b]
\mp \frac{i}{16 \pi} (A_\mu - b_\mu)
(^*\!f_\mu[b] +  ^*\!H_\mu [A,b,0]) \right)
\label{30}
\end{eqnarray}
where $^*\!H_\mu = \varepsilon_{\mu \nu \alpha} H_{\nu \alpha}$.

Condition (\ref{28}) made the ghost term
independent of the bosonic field $A_\mu$. We shall now impose a
second constraint in order to completely  decouple the
auxiliary field $b'_\mu$ from $A_\mu$
in $S_{eff}$. Indeed, consider  the conditions
\beq
\frac{\delta ^2 S_{eff}[b',A]}{\delta A^a_\rho(y)
\delta \,b'^b_\sigma(x)} = 0
\label{31}
\eeq
In terms of the original auxiliary field
 $b_\mu$  these equations read
\beq
\frac{\delta ^2 \tilde{S}[b,A]}{\delta A^a_\rho(y)\delta b^b
_\sigma(x)}  - \frac{\delta ^2 \tilde{S}[b,A]}{\delta b^a_\rho(y)
\delta b^b_\sigma(x)} + \int d^3u \frac{\delta
^2\tilde{S}[b,A]}{\delta b^c_\beta(u)\delta b^b_\sigma (x)} \,\frac
{\delta V^c_\beta(u)} {\delta A^a_\rho(y)} = 0
\label{32}
\eeq
Eqs.(\ref{32}) can be
easily written in terms of  $L$, $H$ and $V$ as a lengthy equation that 
we shall omit here.

The strategy is now as follows: once a given approximate expression for the fermion determinant is considered, 
one should solve eq.(\ref{32}) in order to
determine functionals $V$ in eq.(\ref{18}) and $G$ in eq.(\ref{21}),
 taking also in account the condition (\ref{28}). In particular, if
one considers the $1/m$ expansion for the fermion determinant,
equations (\ref{28}) and (\ref{32}) should determine the form of $V$
and $G$ as a power expansion in $1/m$. In ref.\cite{Red} the $1/m$
expansion for the fermion determinant was shown to give
\beq
\ln \det (\id + m + \bs) =
  \pm \frac{i}{16\pi} S_{CS}[b] +
  I_{PC}[b] +
  O(\partial^2/m^2)  ,
\label{14}
\eeq
where the Chern-Simons action $S_{CS}$ is given by
\beq
  S_{CS}[b] =
 \varepsilon_{\mu\nu\lambda} \tr \int\dv
 (
   f_{\mu \nu} b_{\lambda} -
   \frac{2}{3} b_{\mu}b_{\nu}b_{\lambda}
  )  .
\eeq
Concerning the parity conserving contributions, one has
\beq
I_{PC}[b] =
  - \frac{1}{24\pi  m} \tr\int\dv f^{\mu\nu} f_{\mu\nu}
  + \cdots  ,
\label{8f}
\eeq

To order zero in this expansion, solution of
eqs.(\ref{28}),(\ref{32}) is very simple. Indeed, in this case the
fermion determinant coincides with the CS action and one can easily
see that the solution is given by
\beq
V_\mu^{(0)}[A] = 0
\label{33}
\eeq
\beq
G^{(0)}[A, b, h, \bar c]  =
\pm \frac{i}{16\pi} tr \!\int \! \dv \bar c_\mu \, (
 \frac{1}{2} \, ^*\!f_\mu[A] + \frac{1}{2} \, ^*\!f_\mu[b]
- 2 ^*\!D_{\mu\alpha}[A] h_\alpha
)
\label{34}
\eeq
With this, the change of variables (\ref{18}) takes the simple
form
\beq
b_\mu = 2 b'_\mu - A_\mu
\label{35}
\eeq
and the decoupled effective action reads
\beq
S_{eff}^{(0)}[b,A,\bar c, c] = \mp \frac{i}{16\pi}
(2 S_{CS}[b'] - S_{CS}[A]) + S_{ghost}[b']
\label{36}
\eeq
We then see that the path-integral in eq.(\ref{cor}), defining
the bosonic action $S_{bos}[A]$, factors out so that one
ends with a bosonic action in the form
\beq
S_{bos}^{(0)}[A] = \pm   \frac{i}{16\pi}
 S_{CS}[A]
\label{37}
\eeq
as advanced in \cite{B},\cite{MLNS}. Let us remark that
in finding the solution for $G$ one starts by writing the most
general form compatible with its dimensions,
\begin{eqnarray}
& & G^{(0)}[A, b, h, \bar c] = tr \!\int \! \dv
\varepsilon_{\mu \nu \alpha}
\bar c_\mu \, (
  d_1 b_\nu A_\alpha + d_2 A_\nu b_\alpha + d_3 b_\nu b_\alpha
+ d_4 A_\nu A_\alpha   \nonumber\\
& & + d_5  b_\nu h_\alpha  + d_6  h_\nu b_\alpha
+ d_7  A_\nu h_\alpha + d_8  h_\nu A_\alpha
+ d_9 \partial_\nu A_\alpha + d_{10} \partial_\nu b_\alpha
+ d_{11} \partial_\nu h_\alpha
) \nonumber \\
& & \label{48}
\end{eqnarray}
All the arbitrary parameters $d_i$ are determined by imposing the
conditions (\ref{28}) and (\ref{32}) with $^*\!H_\mu$ transforming
covariantly (as $h_\mu$ does) under gauge transformations which
leads, together 
with a gauge invariant action, to the solution (\ref{34}).

To go further in the $1/m$ expansion one uses the next to the leading
order in the fermion determinant as given in eq.(\ref{14}). Again,
starting from the general form of $G$ and after quite lengthy
calculations that we shall not reproduce here, one can find a unique
solution for $V_\mu$ and $H_{\nu \alpha}$ leading to a gauge
invariant action,
\beq
V_\mu^{(1)}[A] =  \pm \frac{2i}{3m} {^*\!f_\mu[A]}
\label{49}
\eeq
\begin{eqnarray}
& & G^{(1)}[A,b,h,\bar c] = G^{(0)}[A,b,h,\bar c]
 \; \mp \; \frac{1}{96\pi m} tr \!\int \! \dv
\bar c_\mu \varepsilon_{\mu \nu \alpha}
\varepsilon_{\nu \rho \sigma}
\nonumber\\
& &
\left( \; \frac{1}{2} \; [ \; f_{\rho \sigma }[A-h] + 3 f_{\rho
\sigma }[b-h] - 2 D_\rho [A-h] (A_\sigma - b_\sigma) \; , \;
(A_\alpha - b_\alpha) \; ] \right.
\nonumber\\
& &
\left. + \; 4 \; [ \; f_{\rho \sigma }[A-h] \; , \; h_\alpha \; ] \;
\frac{}{} \right)
\label{uff}
\end{eqnarray}
The corresponding change of variables (\ref{18}) takes now the
form
\beq
b_\mu = 2 b'_\mu - A_\mu \pm \frac{2i}{3m} {^*\!f_\mu}[A]
\label{39}
\eeq
and the decoupled effective action reads
\beq
S_{eff}^{(1)}[b,A,\bar c, c] = S_{eff}^{(0)}[b,A,\bar c, c]
+ tr \!\int \! \dv \left(
\frac{1}{6\pi m} f_{\mu\nu}^2[b'] + \frac{1}{24\pi m} f_{\mu\nu}^2[A]
\right)
\label{40}
\eeq
so that one can again  integrate out the completely
decoupled ghosts and $b'$ fields ending with the bosonic action
\beq
S_{bos}^{(1)}[A] =
\pm   \frac{i}{16\pi}
 S_{CS}[A] +  \frac{1}{24\pi m} tr \!\int \! \dv f_{\mu\nu}^2 [A]
\label{50}
\eeq
This result extends to order $1/m$ the bosonization recipe presented
in refs. \cite{B},\cite{MLNS}.

In this way, from the knowledge of the $1/m$ expansion of the fermion
determinant one can systematically find  order by order the
decoupling change of variables and construct the corresponding
action for the bosonic field $A_\mu$. One finds for the change of
variables
\beq
b_\mu = 2 b'_\mu - A_\mu  \pm \frac{2i}{3m} {^*\!f_\mu[A]}
+ \frac{1}{m^2} C^{(2)} D_\rho[A] f_{\mu \rho}[A] + \ldots
\label{51}
\eeq
Here $C^{(2)}$ is a (dimensionless) constant to be determined from the
$1/m^2$ term in the fermion determinant expansion, which should be
proportional to $ {^*\!f_\mu} D_\rho f_{\rho \mu}$. Evidently,
finding the BRST exact form becomes more and more involved and so is
the form of the bosonic action which however, can be compactly
written as
\begin{eqnarray}
& & S_{bos}[A]  =  tr \!\int \! \dv  \left( \frac{}{}
L[-A + V[A]] \right.
\nonumber \\ & &
\left. \mp \frac{i}{16\pi}(2A_\mu - V_\mu[A]) ({^*\!f}_{\mu}[-A +
V[A]] + {^*\!H}_{\mu}[-A + V[A],A,0] ) \right)
\label{52}
\end{eqnarray}
\vspace{0.5cm}

\noindent (ii) {\it The Abelian case}
\vspace{0.5cm}

As it should be expected, the obtention of the bosonic action
from the knowledge of the fermion determinant expansion in powers
of $1/m$ greatly simplifies in the Abelian case. We shall briefly
describe this calculation and then discuss another approximation
which allows to obtain the bosonic action in the $m \to 0$ limit.

No ghosts have to be employed when using the identity (\ref{4})
in the Abelian case. Moreover, since in both approximations
to be  considered only quadratic terms are included in the
determinant expansion, no additional exact BRST terms are necessary
in order to decouple the auxiliary field $b_\mu$ from the bosonic
field $A_\mu$. One then has, instead of (\ref{cor})-(\ref{11x}),
\beq
\exp(-S_{bos}[A])   =  \int \D b_\mu
\exp (-S_{eff}[A,b])
\label{53}
\eeq
with
\begin{equation}
S_{eff}[A,b] =
  \int  \dv ( L[b]
+ i \lambda A_\mu {^*\!\!f}_\mu[b] )
\label{54}
\eeq
(There is no need in the Abelian case  to introduce
an $X[b]$, since the action (\ref{54}) is already
gauge invariant with $A_\mu$ and $b_\mu$ 
transforming as gauge fields.)

Again, we consider the change
of variables 
with
\beq
b_\mu = (1 - \theta) b'_\mu  + \theta A_\mu + V_\mu[A]
\label{55}
\eeq
where $V_\mu[A]$ is a (gauge invariant) functional of $f[A]$ so that
$b'_\mu$ is also a gauge field. Parameters $\theta$ and $\lambda$
are at this stage  arbitrary, and $\theta$ can be a functional of
$\partial ^2/m^2$ (One could have introduced these parameters also 
in the non-Abelian
case but they were chosen so as to simplify calculations
in the $1/m$ expansion). 
In terms of $L[b]$ and $V_\mu[A]$ the decoupling
conditions (\ref{32}) become here
\begin{eqnarray}
& &  2i \lambda \varepsilon_{\rho \sigma
\alpha} \partial_x^\alpha \delta(x-y)
+ \theta _y \int d^3z \frac{\delta^2 L[b(z)]}{\delta
b_\rho(y)\delta b_\sigma(x)}
\nonumber \\
& & + \int d^3u \left( \int d^3z \frac{\delta^2 L[b(z)]}{\delta
b_\beta(u) \delta b_\sigma(x)} \right) \frac{\delta
V_\beta(u)}{\delta A_\rho(y)} = 0 \label{56}
\end{eqnarray}

If we now use for $L[b]$ the result of
the $1/m$ expansion of the fermion determinant \cite{Red},
\beq
L[b] = \mp \frac{i}{8\pi} \varepsilon_{\mu \nu \alpha }b_\mu
\partial _\nu b_\alpha + \frac{1}{24\pi m}f_{\mu \nu }^2[b]
+ O({1\over {m^2}})  \label{57}
\eeq
and try for $V_\mu[A]$
the functional form
\beq
V_\mu[A] = i \frac{C}{m} \varepsilon_{\mu \nu \alpha }
f_{\nu \alpha }[A]
\label{58}
\eeq
we get, from the decoupling equation (\ref{56}),
\beq
C = \pm 1/3
\label{60}
\eeq
if we choose $\lambda = \mp 1/8 \pi$ to have for
simplicity (as in the non-Abelian case)  $\theta = -1$.
The bosonic action for $A_\mu$ can be easily found to be
\beq
S_{bos}[A] = \pm \frac{i}{8 \pi} \int \dv \varepsilon_{\mu
\nu \alpha }A_\mu \partial _\nu A_\alpha  + \frac{1}{24 \pi m}
\int  \dv f_{\mu\nu}^2[A] + O(1/m^2)
\label{61}
\eeq
This result extends that originally presented in ref.\cite{FS}. One
can in principle determine, following the same procedure, the
following terms in the $1/m$ expansion of $S_{bos}$ by including the
corresponding terms in the fermion determinant expansion. It is
evident that, in the absence of an additional BRST exact term, the
decoupling equation (\ref{56}) can have a non-trivial solution provided
one uses for $L[b]$  an approximation  only
involving  terms
quadratic in the fields. This in turn implies that $V_\mu[A]$ should be
a linear functional. Higher order terms in the $1/m$
expansion  involve cubic and higher order terms
in $b_\mu$ thus necessarily requiring  the addition of a BRST
exact term in order attain the decoupling of $b_\mu$ and $A_\mu$.

Alternatively to the $1/m$ determinant expansion, 
one can consider an expansion in powers of $b_\mu$ retaining
up to quadratic terms. The result can be written
in the form \cite{BFO}
\beq
L[b]= \frac{i}{2}\varepsilon _{\mu \nu \alpha }b_\mu
P\partial _\nu b_\alpha  + \frac{1}{  {4 m}}f_{\mu \nu
}[b]Qf_{\mu \nu }[b]
\label{62}
\eeq
where $P$ and $Q$ are functionals to be calculated within a loop
expansion,
\beq
P\equiv P\left( \frac{\partial ^2}{m^2}
\right)\;\;\;\;\;\;\;\;\;\;\;\;\;\;\;\;\;Q\equiv Q\left(
\frac{\partial^2}{m^2}
\right)
\label{63}
\eeq

In order to decouple the $b_\mu$ field one again proposes a change
of variables like in (\ref{55}) but now trying for $V_\mu $
the (gauge-invariant) functional form
\beq
V_\mu [A] =
\frac{i}{ m}\varepsilon _{\mu \nu \alpha }
R f_{\nu \alpha }[A]=2\frac{i}{m}\varepsilon _{\mu \nu \alpha }
R \partial _\nu A_\alpha
\label{64}
\eeq
with
\beq
R \equiv R\left( \frac{\partial ^2}{m^2} \right)
\label{65}
\eeq
One finds, from the decoupling conditions (\ref{56}),
\begin{eqnarray}
&&
\frac{\delta^2 S_{eff}[b',A]}{\delta A_\rho (y)\delta b'_\sigma(x)}
= \nonumber \\
& & (1-\theta )\;\left( \;i \varepsilon _{\rho \sigma \alpha}
\left(
2 \lambda + \theta P - 2 \frac{\partial ^2}{m^2} Q R
\right)
\right. \partial _\alpha \delta (x-y)
\nonumber
\\
& &\left. + \; \frac{2}{m}
\left( \frac{1}{2} \theta Q - P R  \right)
\left( {\partial _\rho \partial _\sigma
-\delta _{\rho \sigma }\partial ^2}\right)\delta (x-y)\;
\right)\;\;\;=0
\label{66}
\end{eqnarray}
$\theta$ being here a functional of $\partial ^2/m^2$.
The solution of this equation is
\beq
R=-\lambda \frac{Q}{\left( {P^2-{\frac{\partial ^2}{m^2}}Q^2}
\right)}\;\;\;\;\;\;\;\;\;\;\;\;\theta =-2\lambda
\frac{P}{\left(
{P^2-{\frac{\partial ^2}{m^2}}Q^2} \right)}
\label{67}
\eeq
With this choice, the change of variables decouples the $b_\mu$
integration so that one can finally get the bosonic action for
$A_\mu$ which now reads
\begin{eqnarray}
 S_{bos}[A] & = & \int \dv \left(
-(2 \lambda )^2\frac{i}{2}\varepsilon _{\mu \nu \alpha }
A_\mu \frac{P}{\left( {P^2-{\frac{\partial ^2}{m^2}}Q^2}
 \right)}\partial
_\nu A_\alpha  \right.
\nonumber \\
& & \left. + (2 \lambda )^2\frac{1}{4 m}
f_{\mu \nu }[A]
\frac{Q}{\left( P^2-\frac{\partial ^2}{m^2} Q^2 \right)}
f_{\mu \nu}[A] \right)
\label{68}
\end{eqnarray}
This result coincides with that found in ref.\cite{BFO},
obtained by a direct functional integration on $b_\mu$.
As it was proven in this last work, it corresponds 
for massless fermions
to the bosonization action  proposed in
ref.\cite{Mar}.

Let us point that in previous works on bosonization based
in the introduction of auxiliary fields $b_\mu$
one needed  the 
explicit (gaussian)
integration on $b_\mu$, this being only possible in the Abelian case 
and in the context of a quadratic approximation. 
An important  point in
the derivation above is that we have shown the equivalence
between our approach to bosonization  and that
based in the direct functional
integration on $b_\mu$. 
Now, in the non-Abelian case any approximation of
the fermion determinant is non-quadratic and a closed functional
integration on $b_\mu$ is not possible, but the approach
presented here can still
be followed. This is due to the fact that 
 an appropriate change of
variable and addition of BRST exact terms
allows to factor out the $b_\mu$ integration, this
being
also true within approximation schemes different 
from the $1/m$ expansion.

In summary, we have developed in the present work a systematic
procedure to construct the bosonic action accompanying the
fermion current bosonization recipe. In the
path integral  approach to bosonization, one
needs 
to decouple an
auxiliary field from the vector field $A_\mu$ describing the system in the bosonic language. We showed that an adequate change of variables
and the existence of a BRST symmetry allows  this decoupling.
In our scheme, one has to determine a functional $V_\mu[A]$ appearing
in the change of variables (eq.(\ref{18})) and an exact BRST term
$\delta G$ to be added to the effective action. We derived two
equations (eqs.(\ref{28}),(\ref{32})) that allow to determine $V_\mu$ and
$G$ and showed through formula  (\ref{52})
how to construct the bosonic
action. Of course one also needs to calculate
the fermion determinant,  
this implying some kind of approximation. For the non-Abelian case we
considered the $1/m$ approximation showing how to construct the bosonic
action order by order in the $1/m$ expansion
(eqs.(\ref{50})-(\ref{51})). We also discussed
in the Abelian case a quadratic approximation for the fermion determinant.
In both cases we obtained consistent formulae for the bosonic
action (eqs.(\ref{61}), (\ref{68})) which, together with eq.(\ref{1}), 
give
the bosonization recipe for three dimensional fermions. Let us
end by noting that the same approach could in principle
be applied in $d>3$ dimensions.

\vspace{0.3 cm}

\underline{Acknowledgements}: The authors wish to thank
 Georges Girardi for helpfull suggestions
and comments. F.A.S. and
C.N.  are partially supported
by Fundacion Antorchas, Argentina and a
Commission of the European Communities
contract No:C11*-CT93-0315.

\end{document}